\documentclass[%
 reprint,
 amsmath,assume,
 aps,floatfix
]{revtex4-2}

\usepackage{amssymb}
\usepackage{titlesec}
\usepackage{hyperref}
\usepackage{float}
\usepackage[section]{placeins}
\usepackage{graphicx}
\usepackage{dcolumn}
\usepackage{bm}

\emergencystretch=\maxdimen
\titleformat*{\section}{\normalfont\bfseries}
\titleformat*{\subsection}{\normalfont}
\titleformat*{\subsubsection}{\normalfont}

\DeclareUnicodeCharacter{2212}{-}
\begin{document}

\preprint{APS/123-QED}

\title{Stretch-induced tunability of electrical transport properties of three-dimensional graphene-based foam structures}

\author{Shu-Ting Guo}
\author{Fangxin Zou}%
 \email{frank.zou@polyu.edu.hk}
\affiliation{%
 Department of Aeronautical and Aviation Engineering, The Hong Kong Polytechnic University, Hung Hom, Kowloon, Hong Kong SAR, China\\
 }%




\date{\today}

\begin{abstract}
The fast electron transport and superior multidirectional flexibility of three-dimensional graphene-based foams (GFs) are pivotal in the realm of stretchable electronics. We observed pre-stretching induced modulation of the temperature-dependent electrical resistivity of GFs, where, as the pre-stretch strain level increased, the distinct temperature dependence of the resistivity of a GF sample would change and might even exhibit a notable transition from negative dependence to positive dependence. We attempted to interpret the phenomenon by proposing a new conduction network model that represents GF structures as interconnected graphene islands and island/island conduction junctions and incorporates three conduction mechanisms: thermally activated conduction, phonon-limited conduction, and fluctuation-induced tunneling conduction. By fitting-assisted analysis, we found that the temperature dependence of the resistivity of a GF sample primarily relies on the discrete quantities of graphene islands and island/island conduction junctions, and the resistivity originating from each conduction mechanism. As pre-stretch strain level increases, these factors would change due to conduction network alteration, local strain-induced phonon hardening, and local strain-induced transport gap modulation, all resulting from pre-stretching. Our results offer valuable insights into the optimization of GFs-based stretchable electronic devices, such as performance enhancement through structural modifications.

\end{abstract}


\maketitle


\section{Introduction}\label{sec1}

In recent years, there has been a growing interest in three-dimensional (3D) graphene-based foams (GFs), which not only retain the unique properties of two-dimensional graphene but also exhibit excellent multidirectional flexibility\cite{huo2019n,javadi2020kinetics,lu2016toward,chen2017lattice,bi2015new,ding2022recent}. These 3D architectures have emerged as promising candidates for the development of stretchable electronic devices, owing to their diverse modes of mechanical deformation, rapid electrical and thermal transport pathways, and easily modulated electrical conductivity\cite{qiu2014mechanically,rong20223d,zhang2022human,ho2019ultralightweight,chen2011three,zheng2018sliced,xu2014facile}. Elastic polymers, such as poly(dimethyl siloxane) (PDMS), have been introduced into GFs as substrates to enhance their mechanical flexibility when used in stretchable electronics, enabling easy twisting, bending, compressing, and stretching of GF/PDMS composite while maintaining good electrical conductivity\cite{chen2011three, pang2016flexible, xu2014facile, jeong2015highly, zheng2018sliced}.

Earlier, we performed an investigation on the electrical resistivity of PDMS-supported 3D GF-based stretchable devices under ramping temperatures ($\rho(T)$). The characterization of the temperature dependence of resistivity was based on the definition of insulating behavior (metallic behavior) as $d\rho/dT$ being negative (positive) \cite{heo2011nonmonotonic}. During this investigation, we happened to observe an intriguing phenomenon where GFs may exhibit either insulating behavior or metallic behavior in temperature dependence, see Figure \ref{fig:1}, and $d\rho/dT$ of GFs could further change with increasing pre-stretch strain levels, see Figure \ref{fig:6} and Figure S3. In some cases, a notable transition from insulating behavior to metallic behavior was observed. The GF structures discussed here are composed of multilayer graphene (MLG) and ultrathin graphite (UG) produced \textit{via} chemical vapor deposition (CVD). Detailed experimental procedures can be found in Methods.

Numerous explorations on the electrical and thermal properties of GFs under diverse environmental conditions have been conducted. For instance, Y. Ito \textit{et al.}\cite{ito2014high} analyzed the transport properties of nanoporous GFs under various magnetic fields and temperatures, while Pettes \textit{et al.}\cite{pettes2012thermal} conducted a study on the temperature-dependent electrical and thermal transport properties of GFs. However, the impact of stretch processing on the temperature dependence of the electrical transport properties of GFs, which we postulate is the key to deciphering our previous observation and crucial for the development of stretchable electronics, remains inadequately scrutinized.

In this work, to interpret the observed phenomenon, a conduction network model for 3D GFs is first proposed to characterize the resistivity, based on a transformation of the foam structure to interconnected two-dimensional graphene-based building blocks. Through leveraging a theoretical understanding of the electrical transport properties of GFs, we apply the model to discern the sources of differences in $d\rho/dT$ for pristine GFs through experimental data fitting. Moreover, by incorporating the influence of stretch-induced structural variations of GFs, obtained \textit{via} Raman spectroscopy, into the relevant physics parameters during fitting, we further attempt to analyze the resistivity changes of different conduction components in GFs under increasing pre-stretch strain levels, so as to interpret the stretch-induced modulation of $d\rho/dT$ for GFs.

\section{Methods}\label{app0}

\subsection{Fabrication of GF samples}

Our samples consist of nickel foam template-directed CVD-grown GFs and elastic PDMS substrates. The nickel foam template must be removed before pre-stretch processing to ensure the stretchability of GFs. To fabricate a conductive PDMS-supported GF sample, first, a GF was thermally annealed under vacuum at 120 $^{\circ}$C for 2 h. Thereafter, a thin layer of PDMS prepolymer, which is a viscous mixture of base/curing agent, was spay-coated onto the GF, leaving the foam edges uncovered. The resulting GF/PDMS composite was baked at 80 $^{\circ}$C for 2 h to reinforce the graphene structure and prevent structural failure. Following this, the GF/PDMS composite was immersed in an HCl (3M) solution at 80 $^{\circ}$C for 48 h, gradually dissolving the nickel foam template through the contact between foam edges and the acid. After cleaning the GF/PDMS composite with deionized water and overnight drying under vacuum at 50 $^{\circ}$C, electrically conductive silver epoxy was applied onto the ends of the composite as electrodes for electrical resistivity measurement. The conductive PDMS-supported GF was finally formed by further infiltrating the GF with PDMS prepolymer, degassing in a vacuum oven for 30 minutes, and then thermally curing at 80 $^{\circ}$C for 2 h.  

\subsection{Electrical resistivity measurement}
The temperature-dependent electrical resistivities of PDMS-supported GF samples were measured in a heating chamber, after subjecting the samples to different pre-stretch strain levels using a tensile testing platform. At each temperature, resistivity is measured three times and the mean value is used in the result analysis.

\subsection{Characterizations}

Structural characteristics of GFs were observed \textit{via} scanning electron microscopy (SEM, JSM-6490, JOEL, Japan), transmission electron microscopy (TEM, Talos F200X, Thermo Fisher Scientific, USA), and optical microscopy (BX43 universal microscope stand, Olympus, USA). Energy dispersive X-ray spectroscopy (EDS) was measured by JSM-6490. Raman spectroscopy was conducted using the Renishaw inVia Raman microscope (UK).

\section{Results and discussion}\label{sec3}

\subsection{Temperature-dependent electrical resistivity of GF structures}

\begin{figure}[htbp]
\includegraphics[width=0.48\textwidth]{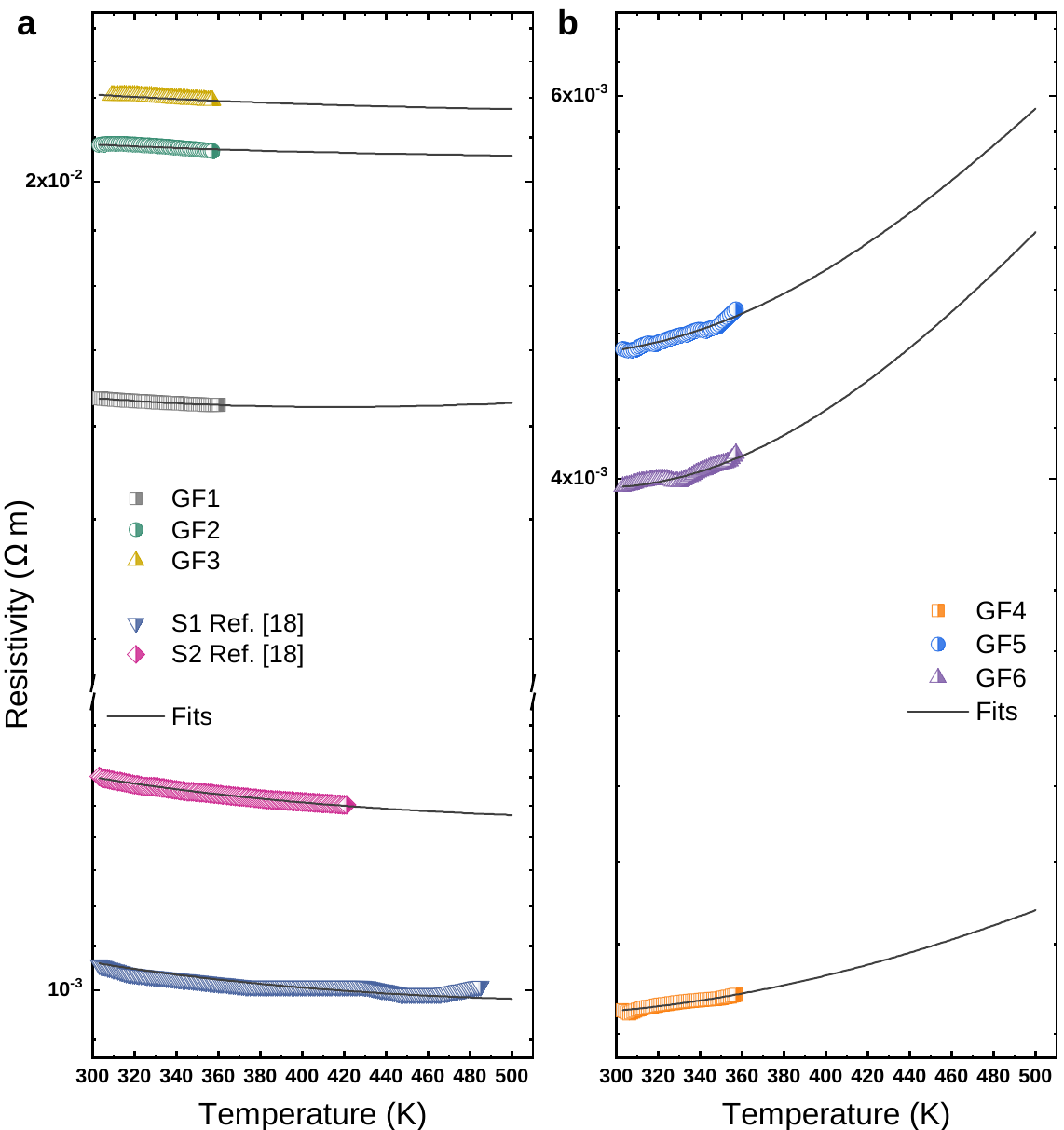}
\caption{\label{fig:1}Temperature-dependent resistivity ($\rho$(T)) results of GF samples. GF1-GF6 are samples investigated in this study, and S1-S2 are samples reported in Ref.18. The scatters are the experimental data. The solid lines are fits of the experimental data to Eq.(\ref{eq:3})-Eq.(\ref{eq:16}) with the parameters listed in Table \ref{tab:table1}. (a) GF1-GF3 and S1-S2 exhibit insulating behavior in temperature dependence ($d\rho/dT<0$). (b) GF4-GF6 exhibit metallic behavior in temperature dependence ($d\rho/dT>0$).}
\end{figure}

The experimentally obtained $\rho(T)$ of representative pristine GF samples are presented in Figure \ref{fig:1}. It's observed that $\rho(T)$ of GF1-GF3 exhibits insulating behavior across the measured temperature range, while $\rho(T)$ of GF4-GF6 shows metallic behavior. A distinct $\rho(T)$ is observed for each sample, aligning with earlier reported studies of bulk and mesoscopic graphite\cite{zoraghi2017influence,garcia2012evidence}. In these studies, the differences in $\rho(T)$ are attributed to the inhomogeneity of graphite material in terms of distributions of crystalline regions and interfaces. Similarly, the temperature-dependent electrical resistivity of two-dimensional CVD-grown MLG/UG is believed to be influenced by the distributions of grain boundaries within each polycrystalline MLG/UG island, as the resistivities across grain boundaries and avoiding grain boundaries within an MLG/UG island exhibit different temperature dependencies\cite{tsen2012tailoring}.

\subsection{Analysis of GF structures}\label{sec30}

\begin{figure}[htbp]
    \centering
    \includegraphics[width=0.49\textwidth]{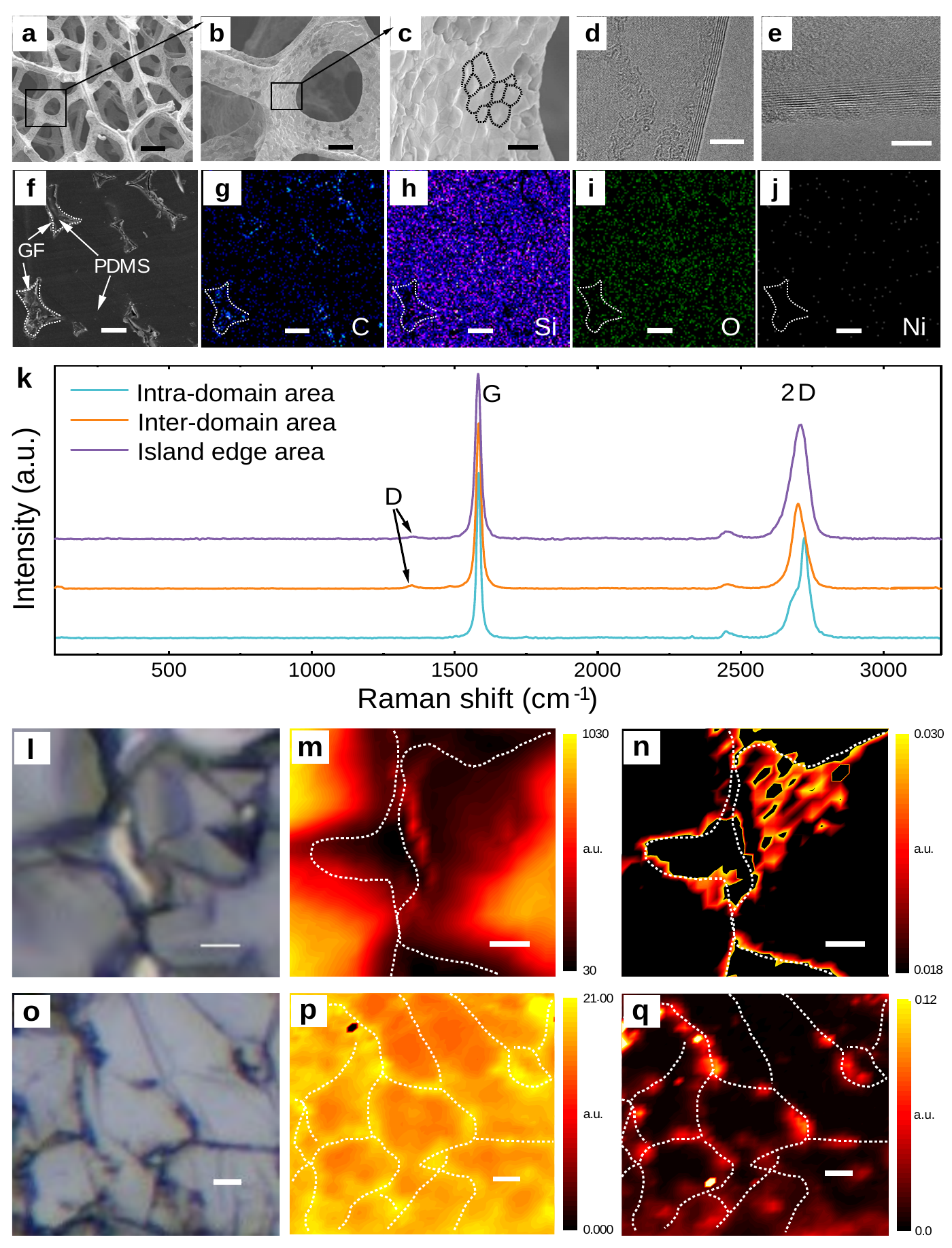}
    \caption{\label{fig:2}(a)--(c) Scanning electron microscope (SEM) observation of GF structures. (a) The porous structure of a GF (scale bar: 200 $\mu m$). (b) Magnified observation of GF strut walls (scale bar: 50 $\mu m$). (c) Magnified observation of strut wall surface (scale bar: 10 $\mu m$). Black dotted lines indicate the locations of domain boundaries within an MLG/UG island. Transmission electron microscope (TEM) observation of (d) MLG with 5 layers (scale bar: 10 $nm$) and (e) UG with 14 layers (scale bar: 5 $nm$) in different areas of the GF strut wall investigated. (f) Cross-section view of a PDMS-supported GF sample. White dotted lines indicate the locations of several GF strut walls. Grey areas are PDMS infiltrated in the GF. (g)--(j) EDS mapping of (f) shows distributions of carbon (C), silicon (Si), oxygen (O), and nickel (Ni), respectively. Scale bars in (f)--(j) are 100 $\mu m$. (k) The 532 $nm$ laser-excited Raman spectra of exampled intra-domain area (avoiding MLG/UG domain boundaries) and inter-domain area (located at MLG/UG domain boundaries) within an MLG/UG island, and edge area of MLG/UG islands. (l) The optical observation of separation between neighboring MLG/UG islands. Raman images of (m) the G band intensity and (n) the D band intensity, which is normalized to the corresponding G band intensity, for the area shown in (l). (o) The optical observation of an MLG/UG island. Raman images of (p) the FWHM of the G band and (q) the D band intensity, which is normalized to the corresponding G band intensity, for the area shown in (o). Scale bars in (l)--(q) are 2 $\mu m$.}
\end{figure}

For exploring the origins of different $\rho(T)$ in the CVD GF samples studied, a comprehensive analysis of GF structures has been conducted. As illustrated in Figure \ref{fig:2}a--c, a 3D porous CVD GF is built by continuous two-dimensional strut walls with various orientations. On the surface of the strut wall, multiple domains divided by domain boundaries can be observed within each MLG/UG island. The observation of GF strut walls, as shown in Figure \ref{fig:2}d,e, confirms the co-existence of MLG and UG throughout the GF samples studied. Additionally, the energy dispersive spectroscopy (EDS) mapping results confirm the removal of the nickel template initially supporting GF strut walls. This is evidenced by the absence of nickel in the strut wall surrounding areas, as depicted in Figure \ref{fig:2}f--j. Accordingly, the impact of nickel on the resistivity of GFs is disregarded in this study, and only GFs composed of MLG and UG are considered.

Further to investigate the distribution of MLG/UG islands throughout an entire GF sample, optical observation is carried out across the sample. The findings suggest that, apart from continuously connected MLG/UG crystalline islands (see Figure \ref{fig:2}o), close contacts between neighboring MLG/UG islands (see Figure \ref{fig:2}l) also contribute to the formation of the GF structures studied. The MLG/UG islands and the contacts between them can be treated as fundamental structural units of GFs, which can be identified through representative peaks in the Raman spectra (see Supplementary Note S1 for detailed descriptions). As shown in Figure \ref{fig:2}k, the prominent G band and broad 2D band obtained in the intra-domain area indicate the existence of multilayer graphene or graphite in the sample\cite{wu2014raman}. The lack of D bands confirms that structural defects are minimal in intra-domain areas. Conversely, in inter-domain areas where domain boundaries are located, and at MLG/UG island edges where structural defects exist, the D band is observed. 

Based on the structural features, the fundamental structural units can be discerned accordingly. As depicted in Figure \ref{fig:2}m,n, the close contact between two MLG/UG islands is reflected by the Raman images of G band intensity and normalized D band intensity, which indicate the location of MLG/UG islands and their edges, respectively. It is postulated that the microstructure of these MLG/UG island contacts is similar to previously reported instances where nanometer-scale gaps are formed between MLG/UG islands due to CVD growth termination\cite{tsen2012tailoring}. The diversely distributed domain boundaries within the MLG/UG island (Figure \ref{fig:2}o) are consistent with the peak values of G band full width at half maximum (FWHM) and normalized D band intensities in the Raman images, as shown in Figure \ref{fig:2}p,q. This suggests the presence of disordered atomic structures in the domain boundaries, which are believed to consist of grain boundaries with topological defects, as reported in previous studies\cite{duong2012probing, kim2011grain, huang2011grains, biro2013grain}. Consequently, we categorize each MLG/UG island as consisting of intra-grain areas (areas avoiding grain boundaries) and inter-grain areas (areas crossing grain boundaries) for further analysis\cite{yu2011control}. The size of graphene grains surrounded by grain boundaries can be estimated from Raman measurement (see Supplementary Note S1 for detailed descriptions).

\subsection{Proposed conduction network model}

The electronic structure of 3D porous graphene has been found to illustrate a linear electronic density of states near the Fermi level, which is consistent with that of two-dimensional graphene\cite{ito2014high}. The model of Lemlich\cite{lemlich1978theory, goodall2006electrical} can therefore be employed to relate the directly measured solid resistivity of foam structures ($\rho$) to the resistivity of two-dimensional strut walls which represents the effective resistivity of foam structures ($\rho_f$). The model has demonstrated its accuracy in describing the electrical transport properties of foam structures compared to other semiempirical models \cite{goodall2006electrical,pettes2012thermal}. Following the model, the effective foam resistivity can be expressed through
\begin{equation}
\rho= \frac{(1-\phi)\rho_{f}}{3}
\end{equation}
where $\phi$ is the porosity of foam structures. To avoid the influence of inaccurately measured GF porosity, here we study $\rho_f$ by normalizing the measured $\rho$ at different $(T)$ temperatures to that at the initial temperature $(T_0)$ as
\begin{equation}
\frac{\rho(T)}{\rho(T_0)}= \frac{\frac{(1-\phi)\rho_{f}(T)}{3}}{\frac{(1-\phi)\rho_{f}(T_0)}{3}}=\frac{\rho_{f}(T)}{\rho_{f}(T_0)}
\end{equation}
where $\rho(T)$ and $\rho(T_0)$ represent the experimentally measured solid resistivities of foam structures, and $\rho_f(T)$ and $\rho_{f}(T_0)$ denote the effective resistivities of foam structures.

\begin{figure}[htbp]
\includegraphics[width=0.48\textwidth]{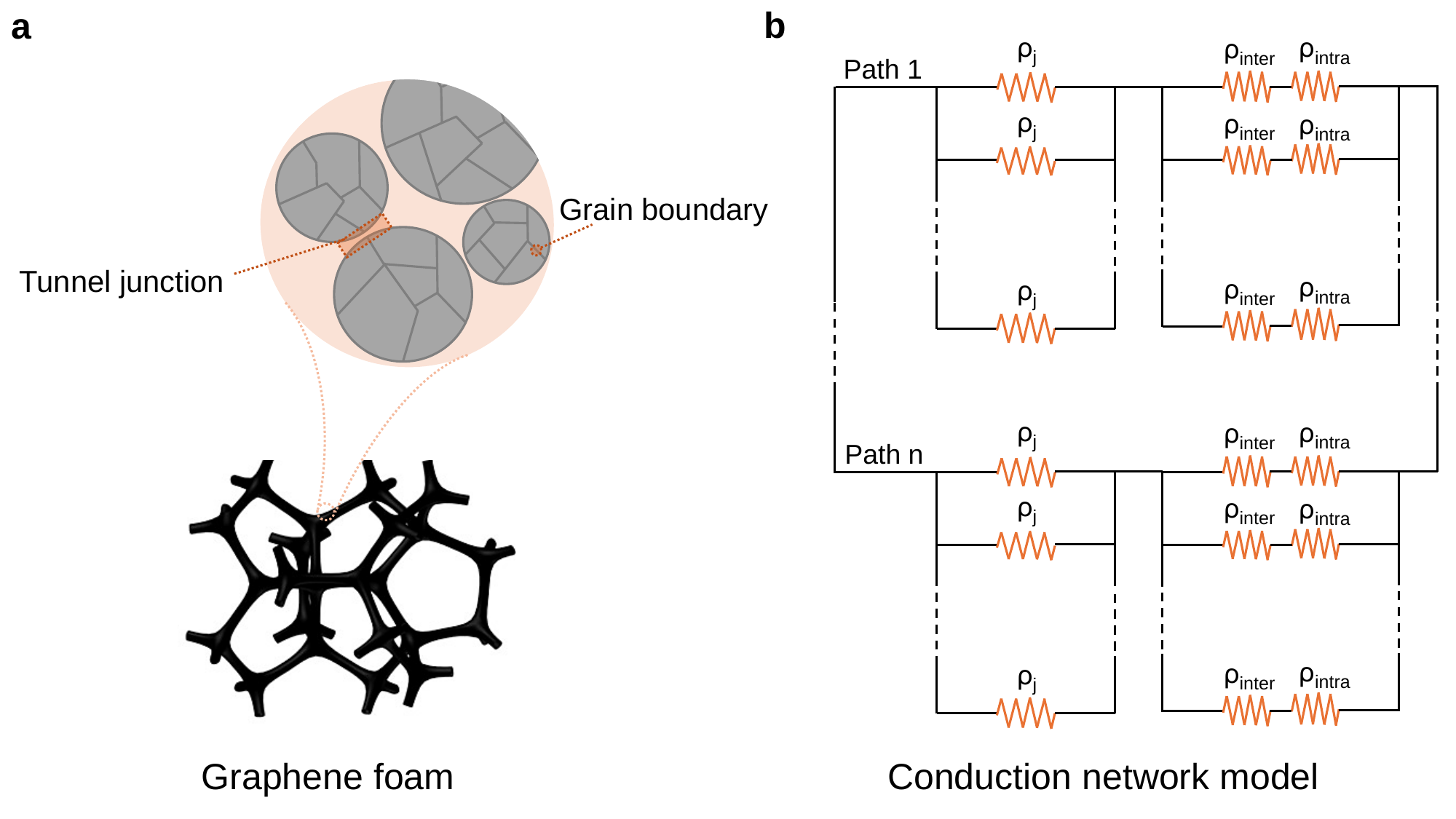}
\caption{\label{fig:b1}(a) Schematic drawing of structural units of graphene foams. (b) A conduction network model demonstrating $n$ parallelly connected effective conduction paths, each incorporating tunnel junction resistivity $\rho_j$ and MLG/UG island resistivity $\rho_d$. Here each $\rho_d$ is further presented by inter-grain resistivity $\rho_{inter}$ and intra-grain resistivity $\rho_{intra}$ in series.}
\end{figure}

The normalized effective foam electrical resistivity can be formulated by considering MLG/UG islands and the contact between neighboring islands (See Figure \ref{fig:2}). The contact between MLG/UG islands can be approximated as a tunnel junction, where the electron tunneling probability is modulated by voltage fluctuation across the junction connecting two conductive segments\cite{sheng1980fluctuation}. The interconnected strut walls are then assumed to be an assembly of randomly distributed MLG/UG islands and tunnel junctions in two dimensions. A conduction network model is therefore proposed for describing the resistivity of GF structures. In the model, as shown in Figure \ref{fig:b1}, the GF conduction system is considered as several parallelly connected effective conduction paths, where each path is formed by a serial resistor network with several equivalent MLG/UG island resistivities in parallel and several equivalent tunnel junction resistivities in parallel. The total resistivity of the conduction system can be formulated as\cite{vzevzelj2012percolating,shim2015optimally,adinehloo2023phonon}:
\begin{equation}\label{eq:3}
\rho_f=N_{path}^{-1}(N_j^{-1}\rho_j+N_d^{-1}\rho_d)
\end{equation}
where $N_d$ is the effective number of MLG/UG islands within each effective conduction path, $N_j$ is the effective number of tunnel junctions within each effective conduction path, and $N_{path}$ is the effective number of conduction paths in parallel in the system. Here, $\rho_d$ and $\rho_j$ are used to represent the resistivity of an equivalent MLG/UG island and an equivalent tunnel junction, respectively. 

For examining the electrical resistivity of MLG/UG islands, we have expanded the conduction network model to incorporate the contribution of transport within both intra-grain areas and inter-grain areas, as discussed in the GF structure analysis. The presence of inter-grain areas, which consist of grain boundaries, can be understood as an extension of the conduction channel that separates the highly conductive intra-grain areas from each other\cite{tsen2012tailoring,song2012origin,vesapuisto2011growth}. As shown in Figure \ref{fig:b1}b, the resistivity of an MLG/UG island with randomly oriented grain boundaries can be considered as contributed by resistors in series and can be calculated based on an Ohmic scaling law through\cite{cummings2014charge,tsen2012tailoring}:
\begin{equation}\label{eq:4}
\rho_d=\rho_{intra}+\frac{\rho_{inter}}{L_d}
\end{equation}
where $\rho_{intra}$ is the average resistivity of intra-grain areas, $\rho_{inter}$ is the average inter-grain resistivity, and $L_d$ is the average grain diameter, which can be estimated through Raman measurements. An average value of $L_d=720\ nm$, see Supplementary Note S1, will be used for further analyzing the electrical transport in the GF samples studied.

In the context of the outlined conduction network model (Eq.(\ref{eq:3}) and Eq.(\ref{eq:4})), the temperature-dependent electrical resistivities of all GF samples demonstrated are interpreted through the fitting of normalized $\rho(T)$ to the model as
\begin{equation}\label{eq:4.1}
\begin{aligned}
\frac{\rho(T)}{\rho(T_0)}&=\frac{\rho_f(T)}{\rho_f(T_0)}\\
&=\frac{N_{path}^{-1}[N_j^{-1}\rho_j(T)+N_d^{-1}\rho_d(T)]}{N_{path}^{-1}[N_j^{-1}\rho_j(T_0)+N_d^{-1}\rho_d(T_0)]}\\
&=\frac{N_j^{-1}\rho_j(T)+N_d^{-1}[\rho_{intra}(T)+\frac{\rho_{inter}(T)}{L_d}]}{N_j^{-1}\rho_j(T_0)+N_d^{-1}[\rho_{intra}(T_0)+\frac{\rho_{inter}(T_0)}{L_d}]}
\end{aligned}
\end{equation}
Details of different terms can be found in the following section.

\subsection{Electrical transport in GF structures}\label{sec31}

\subsubsection{Fitting results}
\begin{figure*}[htbp]
\includegraphics[width=\textwidth]{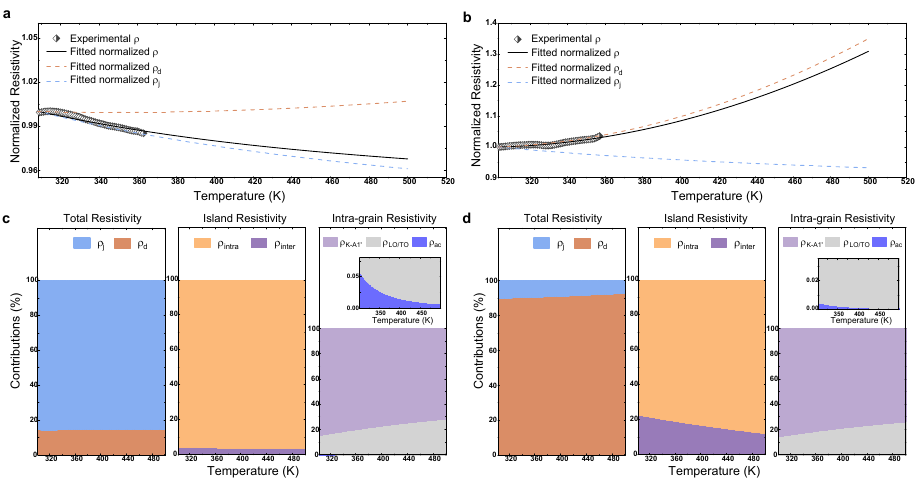}
\caption{\label{fig:4} (a) Normalized temperature-dependent resistivities of GF3. All resistivities are normalized to the corresponding resistivity at the initial temperature of 309K. (b) Normalized temperature-dependent resistivities of GF6. All resistivities are normalized to the corresponding resistivity at the initial temperature of 303K. The experimental $\rho$ are plotted with scattered dots, the fitted normalized total resistivity $\rho$ with continuous black solid lines, and the fitted normalized tunneling junction resistivity $\rho_j$ and MLG/UG island resistivity $\rho_d$ with colored dashed lines. The fitting was conducted using Eq.(\ref{eq:3})-Eq.(\ref{eq:11}) with parameters listed in Table \ref{tab:table1}. The contributions of different conduction components for (c) GF3 and (d) GF6: the contribution of $\rho_j$ and $\rho_d$ in total resistivity, obtained by scaling the resistivity of individual tunnel junction or individual MLG/UG island with its respective quantities $N_j$ or $N_d$; the contribution of intra-grain resistivity $\rho_{intra}$ and inter-grain resistivity $\rho_{inter}$ in island resistivity; and the contribution of acoustic phonon scattering induced resistivity $\rho_{ac}$ (see the inset), optical LO/TO phonon scattering induced resistivity $\rho_{LO/TO}$ and optical $A_{1}^{'}$ phonon scattering induced resistivity $\rho_{K-A_{1}^{'}}$ in intra-grain resistivity.}
\end{figure*}

As observed experimentally, the pristine GF samples studied demonstrate either an insulating or metallic behavior in temperature dependence. By employing the aforementioned conduction network model fitting, good agreement between the fits and experimental data can be achieved for all GF samples studied, as shown in Figure \ref{fig:1}. Here, we present the fitting outcomes for GF3 and GF6 as illustrative examples of distinct temperature-dependent behavior, see Figure \ref{fig:4}a,b. It is evident that, for pristine GF samples, the insulating behavior is primarily influenced by the temperature-dependent resistivity of tunnel junctions, whereas the metallic behavior is predominantly influenced by the temperature-dependent resistivity of MLG/UG islands. Further to explore determining factors of the temperature-dependence in resistivity of GF samples, we proceed to analyze the fitting results by elucidating the transport mechanism of inter-grain conduction, intra-grain conduction, and tunneling junction conduction, and the effectiveness of fitting results for each conduction component. 

\begin{table*}
 \caption{Summary of parameters obtained from the fits of $\rho(T)$ for different GF samples}\label{tab:table1}
 \resizebox{\textwidth}{!}{
    \begin{tabular}{ccccccccccccc}
    \hline   Sample&Tunnel&MLG/UG&Residual&Acoustic&Prefactor&Prefactor&LO/TO&$A_{1}^{'}$&Prefactor&Thermal&Prefactor&Tunneling\\
    &junction&island&&gauge&&&phonon&phonon&&activation&&related\\
    &number&number&resistivity&field&&&Energy&Energy&&energy&&parameter\\
    &($N_j$)&($N_d$)&($\rho_I$)&($\beta_A$)&($\rho_1$)&($\rho_2$)&($E_{op1}$)&($E_{op2}$)&($\rho_a$)&($E_a$)&($\rho_t$)&($T_1/T_0$)\\
     &&&&(eV)&&&(meV)&(meV)&&(meV)&&\\ \hline
     GF1&0.380&0.388&0.147&4.908&3.695&3.007&191.847&157.500&$1.008\times10^{-7}$&15.595&1.689&23.899\\
    GF2&0.230&0.654&5.091&4.922&2.600&1.604&203.256&153.413&$2.703\times10^{-7}$&10.046&6.994&9.124\\
     GF3&0.169&0.939&4.648&4.900&2.500&1.818&210.000&157.500&$1.000\times10^{-7}$&10.000&4.620&12.344\\
     GF4&500.0&0.220&7.177&4.935&12.772&22.891&206.851&143.009&$1.000\times10^{-7}$&10.000&6.528&14.733\\
     GF5&6.289&0.288&1.208&4.956&11.775&11.638&195.415&144.452&$1.000\times10^{-7}$&13.159&0.613&16.245\\
     GF6&2.564&0.303&1.000&4.913&14.288&12.190&192.724&142.500&$1.000\times10^{-7}$&21.173&1.063&23.825\\
     S1&0.040&0.077&0.800&4.929&87.843&80.092&200.139&150.060&$1.000\times10^{-6}$&10.415&62.434&3.388\\
     S2&0.060&0.097&1.237&5.000&86.022&81.095&208.807&157.500&$1.707\times10^{-6}$&10.009&82.110&2.711\\
     \hline
    \end{tabular}}
\end{table*}

\paragraph{Inter-grain resistivity.}

As discussed earlier, the inter-grain areas in MLG/UG islands consist of one-dimensional (1D) grain boundaries, which contribute to the inter-grain resistivity due to electron transport through grain boundaries. Yazyev and Louie\cite{yazyev2010electronic} have theoretically explained the transport based on the momentum conservation principle, considering grain boundaries as intrinsic topological defects of graphene atomic structures. They found either high transparency or perfect reflection of charge carriers from individual grain boundaries over extensively large electron energy ranges. The energy range where charge carriers are backscattered is defined as the transport gap due to the presence of grain boundaries. The conduction of grain boundaries under various temperatures was found to be in accordance with the Arrhenius equation\cite{kumari2014electrical,song2012origin,arrhenius1889dissociationswarme}:
\begin{equation}\label{eq:5}
\rho_{inter}(T)=\rho_a\ exp(\frac{E_a}{k_BT})
\end{equation}
where $\rho_a$ is a pre-exponential factor, \textit{i.e.} resistivity value when $T$ $\rightarrow$ $\infty$, $E_a$ is the thermal activation energy required to surpass the transport gap of grain boundaries, and $k_B$ is the Boltzmann constant. For all GF samples demonstrated in this study, the values of $E_a$ obtained from the fits of $\rho(T)$ to our model consistently fall within the range of $\sim$10 meV to $\sim$21 meV (See Table \ref{tab:table1}). These values are comparable to those reported in earlier studies\cite{kumari2014electrical,minari2006temperature,song2012origin,park2013abnormal}. Moreover, the fitted average inter-grain resistivities for the GF samples studied are found to be similar to each other, as shown in Table \ref{tab:table21}, which aligns with the expected outcomes for GF samples obtained under analogous CVD growth conditions. Note that the fitted inter-grain resistivities in this study are smaller than the reported ones in other research works which typically fall in a range around $1\ k\Omega\ \mu m$. This may be due to the under-estimated grain size obtained from $I_D/I_G$ in Raman spectra, which represents the average distance between defects and may be smaller than the true grain size\cite{pimenta2007studying, cummings2014charge}.

\begin{table}[htbp]
\caption{Summary of inter-grain resistivity obtained from the fits of $\rho(T)$ for all GF samples}\label{tab:table21}
\begin{tabular}{cccc}
\hline
&Sample&Inter-grain resistivity&\\
&&($\rho_{inter}$)($\Omega\ \mu m$)&\\\hline
&GF1&0.183&\\
&GF2&0.397&\\
&GF3&0.147&\\
&GF4&0.147&\\
&GF5&0.166&\\
&GF6&0.225&\\
&S1&1.49&\\
&S2&2.504&\\
\hline
\end{tabular}
\end{table}

\paragraph{Intra-grain resistivity.}
Considering that, before electrical resistivity measurement, all GF samples measured have undergone thermal annealing under vacuum at 120 $^{\circ}$C and are fully infiltrated by PDMS with the nickel template dissolved (see Figure \ref{fig:2}f--j), we assume the impurity atoms in GF samples are removed and the adsorption/desorption of molecules from the air is avoided. Thereby, the influences from charged impurities and molecules adsorption/desorption on GF resistivities are minimized in the study\cite{Chen_2008, article}. Additionally, the absence of structural defects within intra-grain areas has been confirmed by Raman mapping of MLG/UG islands on GF strut walls (see Figure \ref{fig:2}p,q). Therefore, it is reasonable to consider that the phonon-limited electron transport is the major origin of resistivity in intra-grain areas, with minimal restrictions from long-range charged impurity scattering and short-range defect scattering\cite{heo2011nonmonotonic, das2011electronic}. Furthermore, the contribution arising from electron coupling to out-of-plane acoustic and optical phonon modes is disregarded due to its zero value by symmetry\cite{PhysRevB.76.045430,park2014electron,sohier2014phonon}, and the discussion will focus on the in-plane phonon modes. Particularly, for intravalley scattering, near the $\boldsymbol{\Gamma}$=(0,0) point in the first Brillouin zone of graphene, the longitudinal and transverse acoustic phonon modes (LA and TA) and longitudinal and transverse optical phonon modes (LO and TO) are taken into account. In addition to intravalley scattering, the optical $A_{1}^{'}$ intervalley phonon mode which arises close to the Dirac point \textbf{K} is included as well\cite{sohier2014phonon,yan2008phonon,park2014electron}. Theoretical expectations\cite{sohier2014phonon, park2014electron} suggest that at low temperatures (lower than $\sim$270K), the scattering of electrons by acoustic phonons (TA/LA modes) can be approximated as elastic, resulting in linearly temperature-dependent resistivity, while optical phonons do not contribute. As temperature increases, the elastic approximation for acoustic phonons remains valid, and the optical phonons start to participate in increasing the resistivity of graphene. Among optical phonon modes, the optical $A_{1}^{'}$ phonons make a more pronounced contribution compared to LO/TO phonons, owing to their lower energy and stronger coupling with electrons. 

In this study, the temperature range examined exceeds ambient conditions. Consequently, the electron scattering by both acoustic and optical phonons is considered in the analysis. A linear term is adopted to describe the acoustic phonon (LA and TA) scattering-induced resistivity ($\rho_{ac}$) in MLG/UG islands.  Within the Boltzmann transport framework, $\rho_{ac}$ has the simple linear expression\cite{sohier2014phonon, park2014electron}, 
\begin{equation}\label{eq:6}
\rho_{ac}(T)=\frac{2\pi\beta_A^2k_BT}{e^2\hbar\upsilon_F^2\mu_S\upsilon_A^2}
\end{equation}
where $\beta_A$ is the gauge field, which emerges from the changes in the local electronic hopping parameters due to bond length variations and appears in the coupling to both LA and TA phonons, $\upsilon_A$ is the effective sound velocity for the sum of TA and LA contributions, $\mu_S=7.66\ kg\ m^{-2}$ is the two-dimensional mass density of graphene, $\upsilon_F=1\times10^6\ m\ s^{-1}$ is the Fermi velocity of electrons, $e$ is the electronic charge and $\hbar$ is the reduced Planck constant. For the optical phonon scattering induced resistivity, it's suggested that\cite{sohier2014phonon} intrinsic optical phonon scattering is a more significant factor in limiting electronic conduction at high temperatures compared to substrate-dependent sources of scattering. Hence, the potential scattering from low-energy remote phonons of PDMS substrate is excluded from our discussions. Only the scattering originating from optical LO/TO phonons at $\boldsymbol{\Gamma}$ point and $A_{1}^{'}$ phonons at \textbf{K} point are incorporated. The resulting resistivity ($\rho_{op}$) can be approximated by an exponential term considering the increasing phonon numbers with temperature according to the Bose-Einstein function\cite{chen2008intrinsic,kaiser2011electronic}:
\begin{equation}\label{eq:7}
\begin{aligned}
\rho_{op}(T)&=\rho_{LO/TO}(T)+\rho_{K-A_{1}^{'}}(T) \\
&=\frac{\rho_1}{exp^{E_{op1}/k_BT}-1}+\frac{\rho_2}{exp^{E_{op2}/k_BT}-1}
\end{aligned}
\end{equation}
where $\rho_{LO/TO}$ is the resistivity arising from LO/TO phonon scattering, and $\rho_{K-A_{1}^{'}}$ is the resistivity arising from $A_{1}^{'}$ phonon scattering. $\rho_1$ and $\rho_2$ are prefactors. $E_{op1}$ and $E_{op2}$ are the energy of optical LO/TO phonons and optical $A_{1}^{'}$ phonons, respectively. Then the resistivity of intra-grain areas, including the temperature-independent residual resistivity ($\rho_I$), can be calculated through\cite{chen2008intrinsic}:
\begin{equation}\label{eq:8}
\rho_{intra}(T)=\rho_I+\rho_{ac}(T)+\rho_{op}(T)
\end{equation}
where the phonon scattering-induced resistivity at both low and high temperatures are encompassed. 

When applying Eq.(\ref{eq:6}) - Eq.(\ref{eq:8}) to describe the intra-grain resistivity in the proposed conduction network model, we control the fitted values of the gauge field ($\beta_A$), optical LO/TO phonon energy ($E_{op1}$) and $A_{1}^{'}$ phonon energy ($E_{op2}$) around the theoretical values, which are $\beta_A=4.97\ eV$, $E_{op1}=\hbar\omega_{TO}=\hbar\omega_{LO}=200\ meV$, and $E_{op2}=\hbar\omega_{A_{1}^{'}} =150\ meV$\cite{sohier2014phonon}. The final fitted values for these parameters are shown in Table \ref{tab:table1}.  According to the first principle study conducted by Park \textit{et al.}\cite{park2014electron}, high-energy optical and zone-boundary phonons in graphene ($\hbar\omega\geq150\ meV$) account for 50\% of phonon-limited conductivity at ambient temperature and become dominant at higher temperatures. Furthermore, it has been reported that zone-boundary $A_{1}^{'}$ phonons contribute more significantly than optical LO/TO phonons above room temperature\cite{sohier2014phonon}. To validate the consistency of the fitted results with these findings, we have plotted the contributions of each phonon mode in the intra-grain resistivity based on the fits of experimental data for each sample. Using sample GF3 (see Figure \ref{fig:4}c) and GF6 (see Figure \ref{fig:4}d) as illustrations, it's evident that the scattering from optical LO/TO phonons and $A_{1}^{'}$ phonons contributes much more significantly to the intra-grain resistivity compared to the scattering from acoustic phonons. Additionally, the contribution of optical $A_{1}^{'}$ phonons outweighs that of optical LO/TO phonons. Similar results can be found for other GF samples studied, see Supplementary Note S3 for details. The agreement between the fitted results and previous findings helps to ascertain the accuracy of the fitting and the reliability of the proposed conduction network model.

\paragraph{Tunneling junction resistivity.}

Similar to the conduction in other disordered materials formed by conductor-insulator composites, in GFs with MLG/UG islands separated by PDMS layers, electrons can transfer across the insulating PDMS layers between large conductive MLG/UG islands \textit{via} tunneling. The relevant tunnel junctions between MLG/UG islands are usually small in size and subject to thermally activated voltage fluctuations across the junction. By modulating the potential barrier across the insulating layer within the junction, the voltage fluctuations can influence the tunneling probability and introduce a temperature-dependent characteristic to the tunneling conductivity\cite{sheng1980fluctuation}. A wide variety of potential barrier shapes, varying from nearly rectangular to nearly parabolic, can be used to study the effects of voltage fluctuations\cite{simmons1963generalized}. For simplification, a parabolic barrier approximation has been applied to describe tunneling conduction\cite{sheng1980fluctuation}. Within this approximation, the tunnel junction width $w$, the initial rectangular potential barrier height $V_0$, the electronic charge $e$, and the dielectric constant of the vacuum ($\varepsilon_{0}=8.85\times10^{-12}$ $F/m$) and the PDMS insulating layer ($\varepsilon_{r}=2.7$ $F/m$) are related through\cite{sheng1980fluctuation}:
\begin{equation}\label{eq:15}
\lambda=0.795e^2/(4\varepsilon_{r}\varepsilon_{0} wV_0)
\end{equation}
Here, $\lambda$ is a dimensionless parameter that governs the resultant shape of the potential barrier after approximation. It's found that such parabolic barrier approximation is particularly accurate when $\lambda = 0.07$. In this scenario, a closed form for fluctuation-induced tunneling resistivity $\rho_j$ arises as\cite{sheng1980fluctuation}:
\begin{equation}\label{eq:9}
\rho_{j}(T)=\rho_t\ exp(\frac{T_1}{T+T_0})
\end{equation}
where $\rho_t$ is a weakly temperature-dependent resistivity prefactor, and parameters $T_1$ and $T_0$ are defined as
\begin{eqnarray}
T_1=\frac{AV_0^2}{8\pi e^2k_Bw} \label{eq:10}\\
T_0=\frac{T_1}{2\chi w\xi(0)} \label{eq:11}
\end{eqnarray}
$\chi=(2mV_0/\hbar)^{1/2}$ is the tunneling constant, $A$ is the tunneling contact area, and $m$ is the effective carrier mass. $\xi(0)$ is a scaling factor that originates from the Taylor expansion of the transmission coefficient around the Fermi level within the WKB approximation and is equal to $\frac{\pi}{8}$ for the parabolic barrier. ${T}_1$ here characterizes the temperature at which significant thermal excitation across the tunnel barrier is likely to occur, while $T_0$ represents the temperature above which thermal voltage fluctuations due to Johnson noise become relevant\cite{statz2020charge,sheng1980fluctuation}. From Eq.(\ref{eq:15}), Eq.(\ref{eq:10}) and Eq.(\ref{eq:11}), the tunnel junction width $w$ can be derived as
\begin{equation}\label{eq:16}
w=22.714\frac{\varepsilon_r\varepsilon_0\hbar^2}{me^2[\xi(0)]^2}(\frac{T_1}{T_0})^2
\end{equation}
We assume that the maximum tunneling junction width between MLG/UG islands, which allows tunneling conduction to take place, is $100 \ \r{A}$\cite{zare2023effective,li2007correlations}. The parameter ratio $T_1/T_0$ is then constrained within the assumption of $w\textless 100\ \r{A}$, which is applied in the fitting of $\rho(T)$ to the conduction network model. Fitted values of $T_1/T_0$ for all GF samples (approximately 9 to 24), as shown in Table \ref{tab:table1}, are comparable to those obtained from earlier experimental studies on carbon-based materials\cite{Lin_2011,lee2015impact,fugetsu2010graphene}. The differences in fitted values of $T_1/T_0$ among samples may be attributed to the random formation of tunneling junctions between MLG/UG islands during the CVD growth process\cite{cummings2014charge}.

To demonstrate the universality of the conduction network model in 3D GFs, we further applied the model to fit $\rho(T)$ of GF samples S1 and S2 from Ref.18. An average grain size of approximately 595 $nm$ is applied in the fitting\cite{pettes2012thermal}. As depicted in Figure \ref{fig:1}a, excellent consistency between the experimental data and fitted results can be observed for $\rho(T)$ with temperatures up to $\sim$500K. The differences between the samples in Ref.18 and those in this study, in terms of the fitted parameters (see Table \ref{tab:table1}) and inter-grain resistivity (see Table \ref{tab:table21}), may be attributed to a different CVD growth condition in Ref.18. The successful fitting suggests the effectiveness of the proposed conduction network model in interpreting the electrical resistivity of GF-based systems.

\subsubsection{Origins of distinct temperature-dependent resistivity of GF structures}

By utilizing the proposed conduction network model in describing the electrical resistivity, we can calculate the contributions of various conduction components to the overall resistivity. This approach allows for an exploration of the underlying factors contributing to the differences in temperature dependence between different GF samples. As shown in Figure \ref{fig:4}c,d, within each MLG/UG island resistivity, the intra-grain resistivity $\rho_{intra}$ makes a greater contribution than the inter-grain resistivity $\rho_{inter}$. This predominance of $\rho_{intra}$ leads to metallic behavior in MLG/UG island resistivity for both GF3 and GF6. When evaluating the conduction network of GFs in its entirety, wherein the resistivity of each MLG/UG island and each tunnel junction are considered and scaled by their respective quantities, denoted as $N_d$ and $N_j$, distinct scenarios emerge. Specifically, for GF3, the contribution of $\rho_j$, attributable to all tunnel junctions, is more pronounced than that of $\rho_d$, which originates from all MLG/UG islands, across the entire temperature range examined. A resultant insulating behavior in the total resistivity of GF is therefore observed. In contrast, GF6 displays an opposite trend with respect to the contribution of $\rho_j$ and $\rho_d$, leading to metallic behavior in the total resistivity of GF. Comparable analytical evaluations for other GF samples are presented in Supplementary Note S3. 
\begin{table}[htbp]
    \centering
    \caption{Comparison of quantities of MLG/UG islands ($N_d$) and tunnel junctions ($N_j$) of all GF samples studied and their temperature dependence of resistivity}
    \begin{tabular}{ccc}
     \hline
    Sample&$N_d/N_j$&Temperature dependence of resistivity\\
    \hline
    GF1&1.021&Insulating behavior\\
    GF2&2.843&Insulating behavior\\
    GF3&5.556&Insulating behavior\\
    S1&1.925&Insulating behavior\\
    S2&1.617&Insulating behavior\\
    GF4&0.0004&Metallic behavior\\
    GF5&0.046&Metallic behavior\\
    GF6&0.118&Metallic behavior\\
    \hline
    \end{tabular}
    \label{tab:3new}
\end{table}

Note that, for all samples analyzed, $\rho_d$ consistently displays metallic behavior in temperature dependence, while $\rho_j$ exhibits insulating behavior. Therefore, the differing temperature dependence in the total resistivity of pristine GF samples can be attributed mainly to variations in parameters linked to the conduction network, specifically $N_d$ and $N_j$. As summarized in Table \ref{tab:3new}, generally, GF samples with $N_d/N_j>1$ exhibit insulating behavior in temperature dependence, while those with $N_d/N_j<1$ show metallic behavior. The discrete quantities of tunnel junctions and MLG/UG islands inherent in each GF sample are due to the stochastic nature of the CVD growth process, thereby resulting in diverse resistivity temperature dependencies among GF samples. 

\subsection{Stretch-induced tunable electrical transport properties in GF structures}\label{sec32}

\subsubsection{Effect of pre-stretch strain on GF structures}\label{sec321}

\begin{figure}[htbp]
\includegraphics[width=0.45\textwidth]{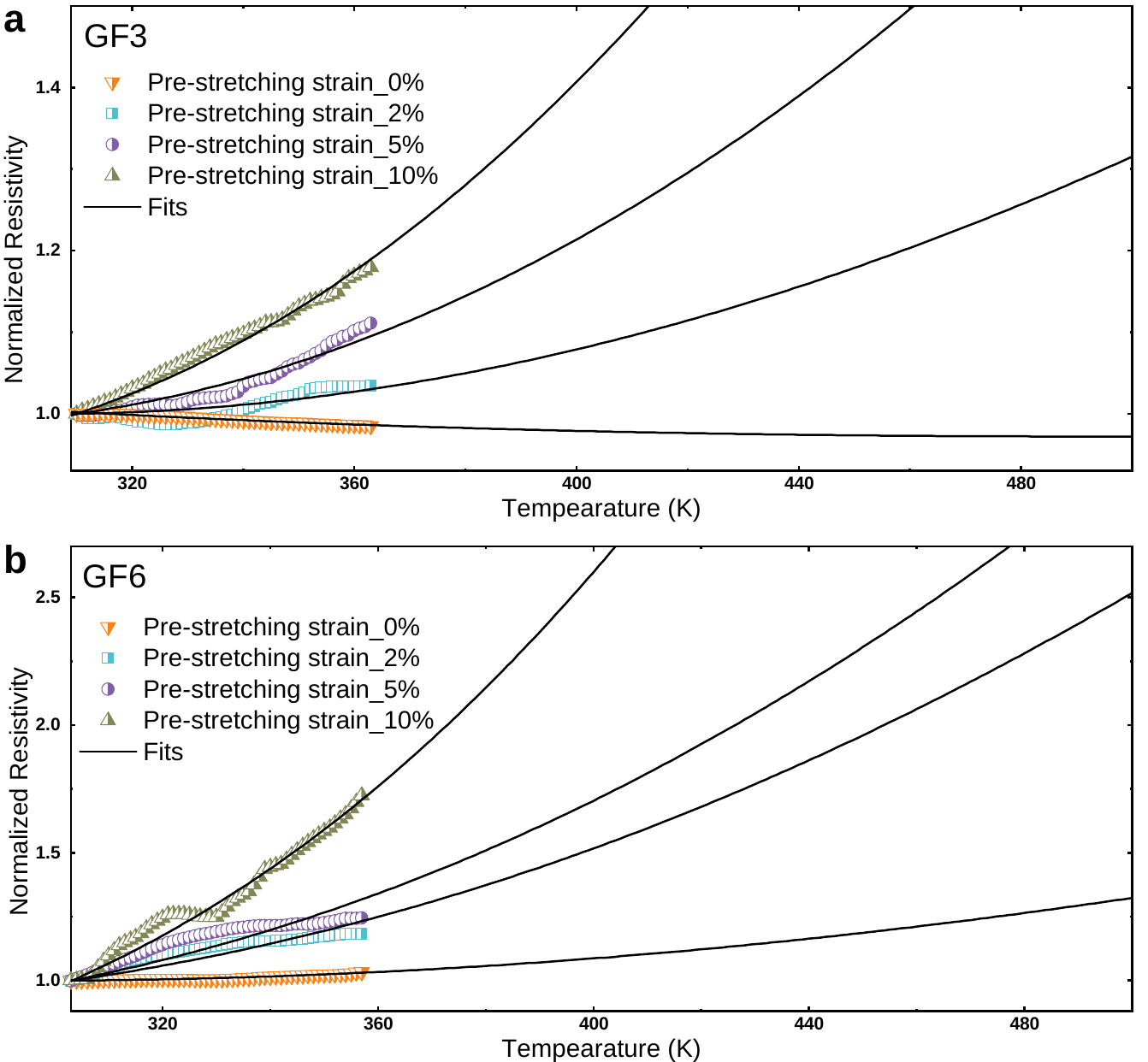}
\caption{\label{fig:6} Normalized $\rho(T)$ of (a) 
 GF3 and (b) GF6 under increasing pre-stretch strain levels. All resistivities $\rho(T)$ here are normalized to the resistivity at the initial temperature. The scatters are the experimental data, and the solid lines are fits of data with parameters in Table \ref{tab:table3}.}
\end{figure}

The demonstrated samples GF1--GF6 were subjected to varying levels of pre-stretch strain, reaching up to 10\%. $\rho(T)$ of each sample was then measured after each pre-stretch strain level. GF3 and GF6 are discussed here as representative samples with an insulating or metallic behavior in temperature dependence. As shown in Figure \ref{fig:6}, increasing pre-stretch strain levels lead to variation in $d\rho/dT$ for both GF3 and GF6. 

To explore the factors influencing stretch-induced changes in $d\rho/dT$, we utilize the approach of conduction network model fitting, taking into account the structural changes in GFs induced by the pre-stretch processing. The conduction network of GF comprises interconnected conduction tunnel junctions and MLG/UG islands with intra-grain and inter-grain areas. In our analysis, for simplicity, we assume that the tunneling contact area, tunnel junction width, and potential barrier height of a fitted equivalent tunnel junction remain unchanged under varying pre-stretch strain levels. Therefore the values of $\rho_t$ and $T_1/T_0$ at the pristine condition for each sample are used in the fitting of corresponding experimental data under different pre-stretch strain levels. Meanwhile, changes in  MLG/UG islands induced by pre-stretching are examined through Raman spectra. The G band frequency and the 2D band frequency after different pre-stretch strain levels are obtained from two other independent GF samples, with a large number of sampling points chosen from intra-grain and inter-grain areas throughout these samples. As shown by Figure \ref{fig:8}a,b, a slight blue shift of the G band and the 2D band is observed with increasing pre-stretch strain levels for both intra-grain and inter-grain areas. A similar phenomenon has been reported for two-dimensional graphene sheets on elastic substrate when uniaxial tensile loading is released\cite{ni2008uniaxial, zeng2020dynamically, srivastava2011raman, bissett2012effect}. For a PDMS-supported graphene sheet, the blue shift of the G band or 2D band can be attributed to the high mobility of PDMS molecular chains which draw in the direction of uniaxial stress leading to lateral compression of the graphene sheet\cite{srivastava2011raman, bissett2012effect}. Additionally, the negative Poisson’s ratio of graphene could also contribute to the compression\cite{woo2016poisson,jiang2016intrinsic}. The frequency shift of graphene Raman feature bands is associated with the uniaxial stress-induced hydrostatic strain and shear strain. In this analysis, as the shear strain has a much smaller contribution, we consider that the strain comes from hydrostatic strain only\cite{reich2000shear}. The Raman shift is therefore converted to hydrostatic strain (See Supplementary Note S2 for detailed conversion procedures). As shown in Figure \ref{fig:8}c, an increasing compressive strain of up to approximately 0.4\% is observed for both intra-grain and inter-grain areas as the pre-stretch strain level increases from 0\% to 10\%. It's theoretically predicted that the G band will not split with a uniaxial strain less than 1\%\cite{cheng2011gruneisen}, which is consistent with the findings in this work (see Figure \ref{fig:8}a,b). The pre-stretching induced hydrostatic strain in both intra-grain and inter-grain areas will be referred to as local strain in the subsequent discussions. The impact of local strain on resistivity in intra-grain and inter-grain areas will be addressed in the experimental data fitting.

\subsubsection{Model parameter changes under pre-stretch strain and fitting results}

\begin{figure*}[htbp]
\includegraphics[width=0.8\textwidth]{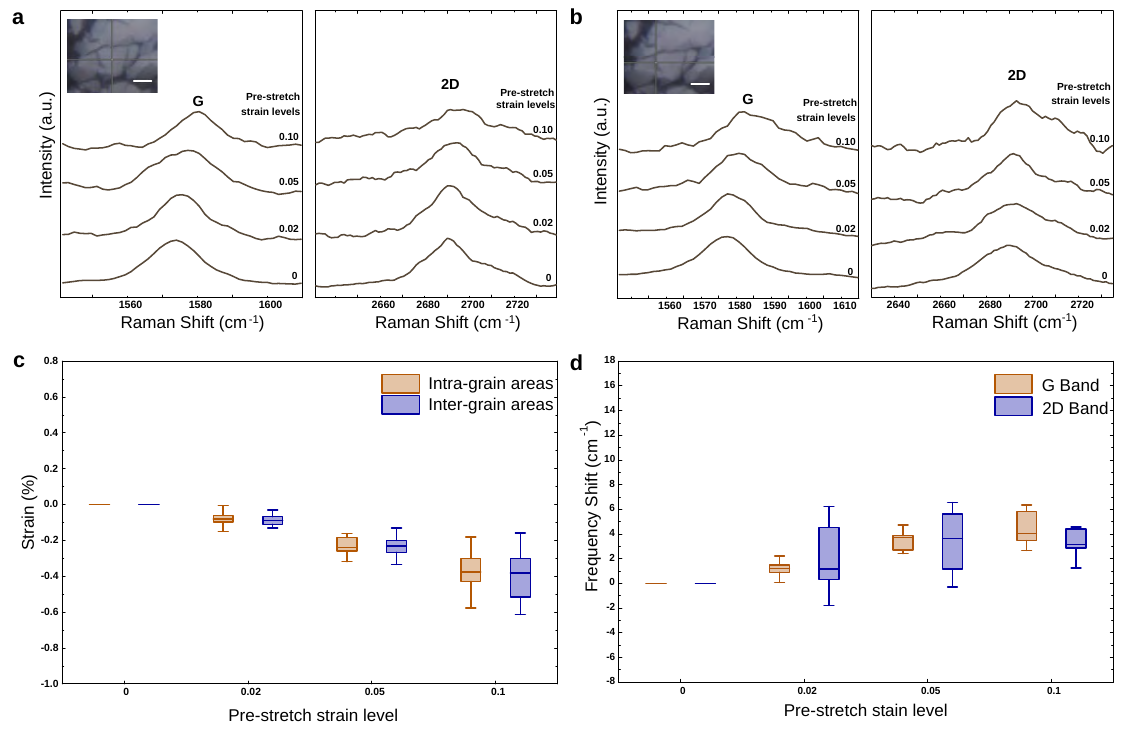}
\caption{\label{fig:8} Changes in the Raman shift of graphene featured G band and 2D band in (a) an intra-grain area and (b) an inter-grain area with increasing pre-stretch strain levels. The insets in (a) and (b) are optical images with crosses denoting the specific testing points for capturing Raman spectra. The scale bars for the optical images are 5 $\mu m$. Box plots of (c) pre-stretching produced local strain and (d) the frequency shift of G band and 2D band for intra-grain areas, based on Raman spectra captured from a number of testing points under different pre-stretch strain levels. The median value of the data group for each pre-stretch strain level is used in the discussion.}
\end{figure*}

\paragraph{Changes in intra-grain resistivity.}

\begin{table*}[htbp]
\caption{Summary of median values of graphene featured G band and 2D band frequency shifts under different pre-stretch strain levels and corresponding energy changes of optical phonons}\label{tab:table2}
\begin{tabular}{ccccc}
\hline
Pre-stretch&G band&2D band&LO/TO phonon energy changes&$A_{1}^{'}$ phonon energy changes\\
strain&$\bigtriangleup \lambda^{-1}$&$\bigtriangleup \lambda^{-1}$&$\bigtriangleup E_{op1}$&$\bigtriangleup E_{op2}$\\
level&$(cm^{-1})$&$(cm^{-1})$&$(meV)$&$(meV)$\\ \hline
 0\% &0&0&0&0\\
 2\% &1.200&1.176&0.152&0.167\\
 5\% &3.703&3.624&0.459&0.403\\
 10\% &4.072&3.168&0.552&0.393\\
 \hline
\end{tabular}\\
\textsuperscript{\emph{a}}Inverse wavelength shift $\bigtriangleup \lambda^{-1}$ can be converted to frequency shift $\bigtriangleup\omega$ \textit{via} $\bigtriangleup\lambda^{-1}=\bigtriangleup\omega/(2\pi c)$.\\
\textsuperscript{\emph{b}}The phonon energy change can be obtained through $\bigtriangleup E=\hbar\bigtriangleup\omega$.
\end{table*}

\begin{table*}
\caption{Summary of parameters obtained from the fits of $\rho(T)$ for representative sample GF3 and GF6 under increasing pre-stretch strain levels.}\label{tab:table3}
\resizebox{\textwidth}{!}{
\begin{tabular}{cccccccccccccc}
 \hline
 Sample&Pre-stretch&Tunnel&MLG/UG&Residual&Acoustic&Prefactor&Prefactor&LO/TO&$A_{1}^{'}$&Prefactor&Thermal&Prefactor&Tunneling\\
    &strain&junction&island&&gauge&&&phonon&phonon&&activation&&related\\
    &level&number&number&resistivity&field&&&Energy&Energy&&energy&&parameter\\
    &&($N_j$)&($N_d$)&($\rho_I$)&($\beta_A$)&($\rho_1$)&($\rho_2$)&($E_{op1}$)&($E_{op2}$)&($\rho_a$)&($E_a$)&($\rho_t$)&($T_1/T_0$)\\
     &&&&&(eV)&&&(meV)&(meV)&&(meV)&&\\ \hline
GF3&0\%&0.169&0.939&4.648&4.900&2.500&1.818&210.000&157.500&$1.000\times10^{-7}$&10.000&4.620&12.344\\
&2\%&100.0&27.027&0.150&4.900&45.090&64.360&209.995&157.5&$1.106\times10^{-6}$&11.018&4.620&12.344\\
&5\%&100.0&76.923&0.100&4.900&42.359&253.420&210.001&157.513&$1.336\times10^{-6}$&12.597&4.620&12.344\\ 
&10\%&1000.0&13.158&0.200&4.900&22.079&30.000&210.000&157.500&$1.000\times10^{-7}$&14.264&4.620&12.344\\ \hline
GF6&0\%&2.564&0.303&1.000&4.913&14.288&12.190&192.724&142.500&$1.000\times10^{-7}$&21.173&1.063&23.825\\
&2\%&66.667&12.500&3.783&4.900&29.452&226.885&192.715&142.513&$1.125\times10^{-7}$&19.918&1.063&23.825\\
&5\%&62.500&9.259&0.700&4.914&20.000&100.790&192.705&142.486&$1.372\times10^{-7}$&17.967&1.063&23.825\\
&10\%&100.0&9.524&0.319&4.913&21.270&196.221&192.744&142.514&$1.000\times10^{-7}$&15.900&1.063&23.825\\
\hline
\end{tabular}}
\end{table*}

Figure \ref{fig:4} suggests that the acoustic phonon scattering-induced resistivity, denoted as $\rho_{ac}$, is minimal. Consequently, $\rho_{ac}$ is assumed to remain unchanged in the presence of pre-stretch strain during the fitting process. The alteration in the scattering of optical LO/TO phonons and $A_{1}^{'}$ phonons induced by local strain is considered the primary factor causing resistivity variation in intra-grain areas. Note that the phonon dispersion of graphene plays a significant role in the description of its Raman spectrum.  The two degenerate in-plane optical LO/TO phonons at $\boldsymbol{\Gamma}$ and the optical $A_{1}^{'}$ phonons of the TO branch near \textbf{K} are the phonons that contribute the most to a graphene Raman spectrum. Specifically, the one-phonon Raman process-induced G band is contributed by the optical LO/TO phonons, and the two-phonon Raman process-induced 2D band originates from two optical $A_{1}^{'}$ phonons with opposite wave vectors\cite{reichardt2017raman, bendiab2018unravelling}. The Raman shift, which corresponds to the energy difference between the incident and scattered photons in a Raman process, of the prominent G band and 2D band in the graphene Raman spectrum can be converted to energy of the optical LO/TO phonon and optical $A_{1}^{'}$ phonons involved, respectively\cite{reichardt2017raman}. The local strain-induced energy alterations of optical LO/TO phonons and $A_{1}^{'}$ phonons can then be tracked by the frequency shift of the G band and 2D band in the Raman spectrum. The Raman spectra collected from different points on the samples show a peak frequency shift, as summarized in Figure \ref{fig:8}d. The upshifts of the G band and 2D band suggest phonon hardening caused by increasing pre-stretch strain levels. The median value of all measured frequency shifts under each pre-stretch strain level is converted to energy and added to the fitted $E_{op1}$ and $E_{op2}$ for each sample prior to pre-stretching. The adjustment accounts for changes in intra-grain area resistivities in the experimental data fitting. The peak frequency shift extracted from Raman measurement and corresponding phonon energy changes used in the data fitting, for each pre-stretch strain level for all samples, are listed in Table \ref{tab:table2}.

\begin{figure*}[htbp]
\includegraphics[width=\textwidth]{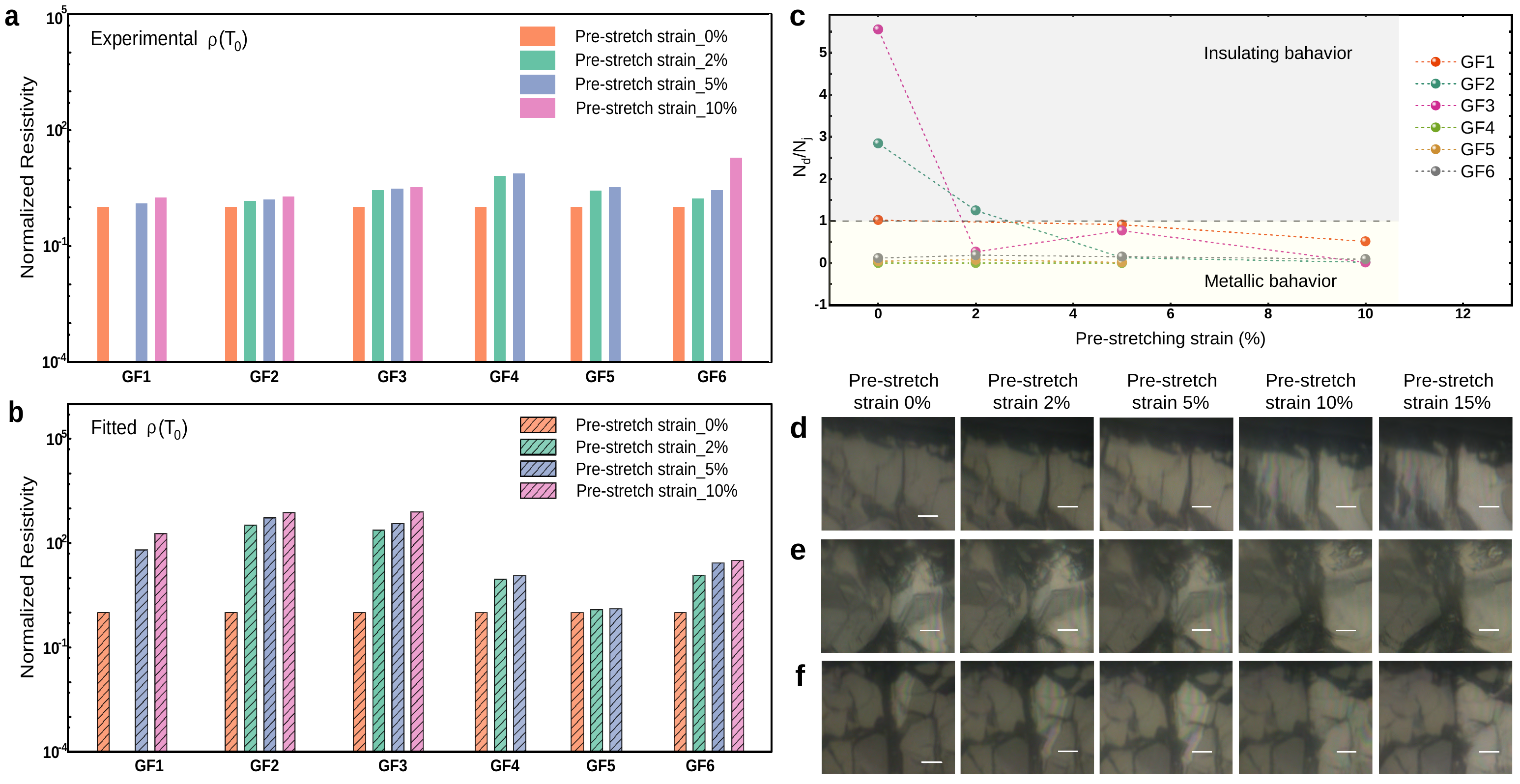}
\caption{\label{fig:9} (a) The experimentally obtained initial resistivity $\rho(T_0)$ and (b) conduction network model fitted $\rho(T_0)$ of GF samples after various pre-stretch strain levels. Each $\rho(T_0)$ is normalized to the corresponding $\rho(T_0)$ at 0\% pre-stretch strain. (c) Summary of $N_d/N_j$ obtained from fitting of experimental data for GF1-GF6 with increasing pre-stretch strain levels. (d)--(f) Examples of optically observed changes in inter-connected MLG/UG island networks with increasing pre-stretch strain levels: (d) and (e) gradual breakage of a complete MLG/UG island, (f) new connections gradually formed by separated MLG/UG islands. Scale bars in (d)--(f) are 5 $\mu m$.}
\end{figure*}

\paragraph{Changes in inter-grain resistivity.}
As previously discussed, the electrical transport between graphene domains is primarily constrained by the transport gap created by grain boundaries. The modulation of the inter-grain resistivity due to pre-stretching can be attributed to the local strain-induced alteration of the transport gap energy. The theory of charge carrier transmission through grain boundaries was developed based on the momentum conservation principle to evaluate the transport gap between two graphene domains with different crystallographic orientations\cite{yazyev2010electronic}. The linear dispersion expression $E(\boldsymbol{k})=\hbar\upsilon_{F}|\boldsymbol{k}|$ is used to characterize the low-energy charge carriers in graphene, where momentum $\boldsymbol{k}$ is measured relatively to the Dirac point in the hexagonal two-dimensional Brillouin zone. When crossing a grain boundary, which is considered a 1D interface with periodic length $d$, the charge carriers experience an effective rotation of the Brillouin zone from the left graphene domain to the right graphene domain. According to translational symmetry, elastic transmission implies that both energy $E$ and momentum $k_{||}$ parallel to the 1D interface are conserved. The correspondence of momentum $\boldsymbol{k}$ in the two-dimensional Brillouin zone of graphene to $k_{||}\in [-\pi/d, \pi/d)$, which indicates the 1D mini-Brillouin zone of a grain boundary, is made by folding two-dimensional momentum space along the lines $k_{||}=(2n+1)\pi/d(n\in \mathbb{Z})$ and projecting the states onto the 1D mini-Brillouin zone. Here, $k_L$ and $k_R$ are utilized to represent the projected momentum $\boldsymbol{k}$ of the left graphene domain and the right graphene domain, respectively. Then the transport gap energy $E_a$ can be approximated as follows\cite{yazyev2010electronic, kumar2012strain}:
\begin{equation}\label{eq:14}
    E_a=\hbar\upsilon_{F}|k_L-k_R|
\end{equation}
where the reduced Plank constant $\hbar=6.582\times10^{-16} eV\cdot s$. In general, the responses of the left and right graphene domains with different orientations, to a uniaxial strain, differ from each other. The modulation of the transport gap by strain can be studied by analyzing the strain effect on the individual graphene domains. As indicated by Eq.(\ref{eq:14}), the changes of $k_L$ and $k_R$ lead to alterations of the transport gap. An analytical expression for momentum $\boldsymbol{k}$ in the two-dimensional Brillouin zone of graphene has been derived by Kummar and Guo\cite{kumar2012strain}, as detailed in Supplementary Note S4. This expression describes the relationship between momentum $\boldsymbol{k}$ and the uniaxial strain perpendicular to the grain boundary direction, taking into account the crystallographic orientations of the graphene domain concerned. The strain-modulated momentum $k_L$ and $k_R$ and the resulting transport gap energy can then be calculated through the change in the corresponding momentum $\boldsymbol{k}$ obtained by this analytical expression. In this work, the $E_a$ values of different GF samples under different pre-stretch strain levels are evaluated following this approach based on the measured local strain in the inter-grain areas as shown in Figure \ref{fig:8}c. Note that the grain boundaries in graphene typically exhibit periodic length $d$ of 1--5 nm, as demonstrated by scanning probe microscopy studies\cite{yazyev2010electronic}. The calculation of $E_a$ is conducted within this observed range of $d$. Knowing the fitted value of $E_a$ of each GF sample before pre-stretching, it is possible to identify an equivalent grain boundary representing the inter-grain graphene structure with a suitable crystallographic orientation angle and periodic length $d$ based on the analytical expression of momentum $\boldsymbol{k}$. The modulated $E_a$ under certain pre-stretch strain levels can therefore be calculated using the identified crystallographic orientation angle and grain boundary periodic length $d$ with the obtained local strain. 

The experimental data of GF samples with insulating behavior or metallic behavior are fitted into the proposed conduction network model considering the influence of local strain on intra-grain resistivity and inter-grain resistivity. The parameters $N_j$ and $N_d$ are allowed to vary during the fitting process to accommodate changes in the conduction network structure.  The experimental data for GF3 and GF6 (see Figure \ref{fig:6}) show good agreement with the fits, and the corresponding parameters are listed in Table \ref{tab:table3}. Successful fitting results for other samples can be seen in Supplementary Note S3.

The comparison between the experimentally measured initial resistivity $\rho(T_0)$ and the fitted ones after each pre-stretch strain level can provide evidence of the effectiveness of the fitting process. As illustrated in Figure \ref{fig:9}a,b, experimentally measured $\rho(T_0)$ consistently increases with higher pre-stretch strain levels for all GF samples. This trend is also observed in the fitted $\rho(T_0)$ values for all samples, suggesting the reliability and accuracy of the conduction network model fitting.

We have also plotted the ratio of parameters $N_d/N_j$ across all samples, see Figure \ref{fig:9}c, to interpret the conduction network changes under increasing pre-stretch strain. For GF1--GF3, the values of $N_d$ and $N_j$ are comparable (see Table \ref{tab:table1}) with a ratio of $N_d/N_j>1$, and insulating behavior is exhibited prior to pre-stretching. This ratio decreases following pre-stretching, indicating a shift of the temperature dependence in resistivity towards metallic behavior. This shift implies that pre-stretching likely causes a more pronounced reduction in $N_d$ compared to $N_j$. This may have resulted from the breakage of intact MLG/UG islands and the simultaneous formation of conduction junctions, which is consistent with the optically observed changes in conduction networks. As shown in Figure \ref{fig:9}d,e, when an MLG/UG island undergoes gradual separation, tunneling conduction junctions may form between the two halves if they remain sufficiently close, as observed at a pre-stretch strain of 2\%. However, increasing the pre-stretch strain to 5\% can lead to the destruction of these junctions, causing a possible increase in $N_d/N_j$, see GF3. On the other hand, at pre-stretch strain over 10\%, new connections can emerge between certain neighboring MLG/UG islands, as demonstrated in Figure \ref{fig:9}f. This would result in the formation of tunneling conduction junctions, an increase in $N_j$, and a corresponding decrease in $N_d/N_j$. For GF4--GF6, which show metallic behavior before pre-stretching, the value of $N_d$ is significantly smaller than $N_j$ (see Table \ref{tab:table1}), and as such, the $N_d/N_j$ ratio exhibits little fluctuations with increasing pre-stretch strain, though the observed conduction network changes shown in Figure \ref{fig:9}d--f also apply to those samples. In summary, the alignment of $N_d/N_j$ ratio variations with the observed changes in the conduction network upon pre-stretching lends further credence to the accuracy of our fitting results.

\begin{figure}[htbp]
    \centering
    \includegraphics[width=0.45\textwidth]{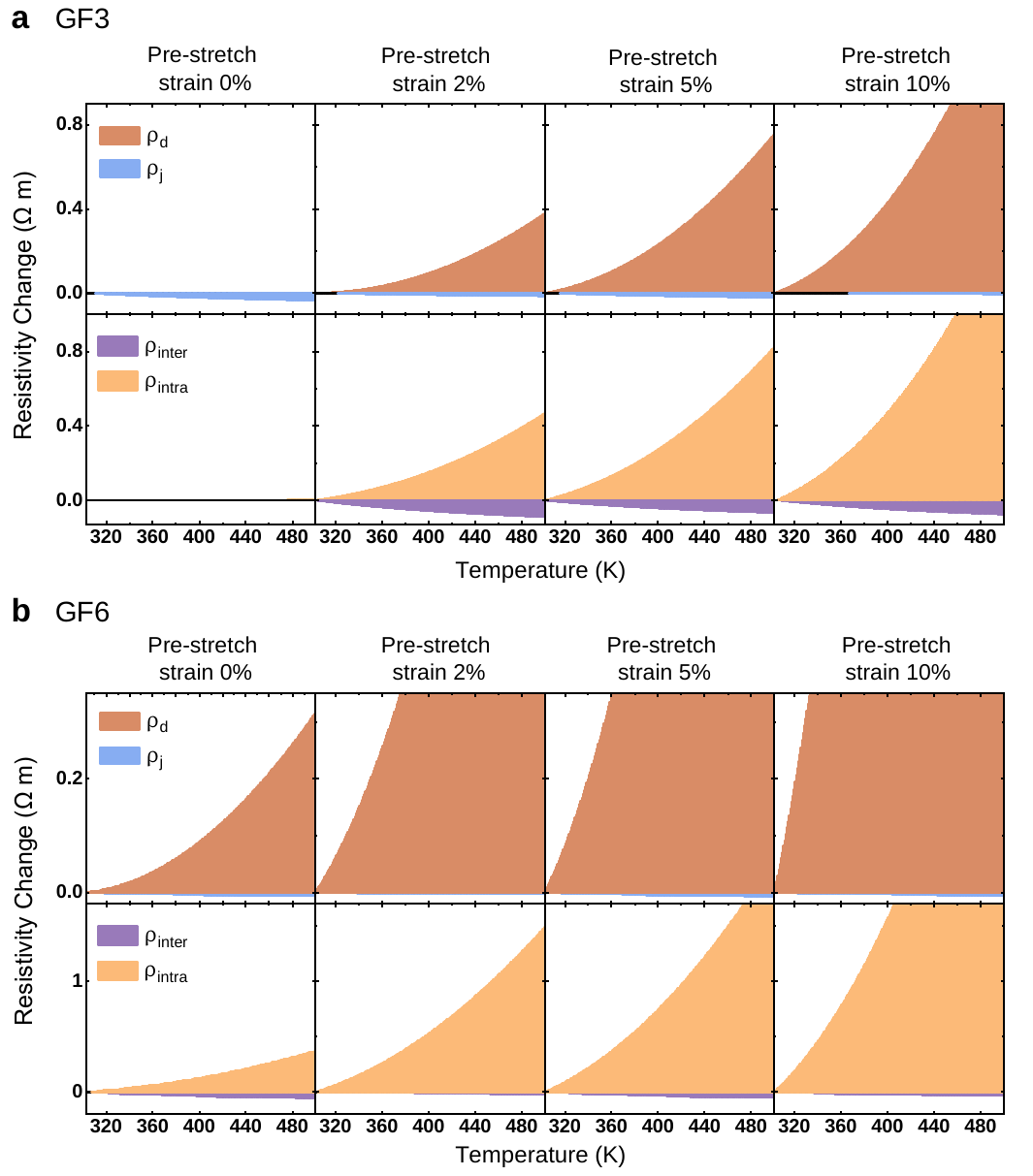}
    \caption{Resistivity change of each conduction component for (a) GF3 and (b) GF6, including tunnel junction resistivity $\rho_j$, MLG/UG island resistivity $\rho_d$, inter-grain resistivity $\rho_{inter}$, and intra-grain resistivity $\rho_{intra}$. The resistivity change is derived from the fitting results by scaling the individual resistivity of each conduction component with the corresponding conduction network-related parameters, specifically, $N_j$ and $N_d$.}
    \label{fig:enter-label2}
\end{figure}

\subsubsection{Origins of pre-stretching induced modulation of temperature-dependent electrical resistivity of GFs}

The primary factor influencing the changes in $d\rho/dT$ under varying pre-stretch strain levels can be identified based on the fitting results. This involves analyzing the resistivity change of each conduction component, including resistivity $\rho_j$, $\rho_d$, $\rho_{inter}$, and $\rho_{intra}$. Here, to account for the entire conduction network of each GF sample, the resistivity changes assessed are determined by scaling the fitted individual resistivity of each conduction component with the corresponding fitted conduction network-related parameters, namely $N_j$ and $N_d$. For GF3, the resistivity change of $\rho_j$ shows a dominant role prior to the pre-stretch processing, as shown in Figure \ref{fig:enter-label2}a, resulting in a $\rho_j$ dominated temperature dependence of total resistivity, which exhibits insulating behavior as previously discussed. With an increase in the pre-stretch strain level, the resistivity change of $\rho_d$, which shows metallic behavior in temperature dependence, is increasingly dominant, leading to a monotonically increasing trend of $d\rho/dT>0$, as shown in Figure \ref{fig:6}a.

Specifically, within the MLG/UG island resistivity change, the dominant role is played by $\rho_{intra}$. The increasing resistivity change of $\rho_d$ could be attributed to the amplified resistivity change of $\rho_{intra}$, which is a result of the combined effect of structure alterations and local strain-induced phonon hardening, both resulting from pre-stretching, as previously discussed. A similar case can be observed for GF6 as depicted in Figure \ref{fig:enter-label2}b. A growing resistivity change is observed for both $\rho_d$ dominated total resistivity and $\rho_{intra}$ dominated island resistivity with increasing pre-stretch strain levels due to the pre-stretching produced local strain and conduction network changes. As a consequence, $d\rho/dT$ exhibits a monotonically increasing trend, as presented in Figure \ref{fig:6}b. Further results analysis of other samples can be found in Supplementary Note S3. It is noteworthy that in some instances, see Figure S3e, a nonmonotonic alteration in $d\rho/dT$ can also be observed, which is attributed to the dynamic changes of the conduction network induced by the pre-stretch processing.

\section{Conclusion}\label{sec4}

We have investigated the temperature-dependent electrical resistivity $\rho(T)$ of GFs and its variations after being subjected to different pre-stretch strain levels. Our findings indicate that the temperature-dependent electrical resistivity of GFs can exhibit either insulating behavior or metallic behavior, and these properties can be modulated by pre-stretch processing. A conduction network model is proposed in this work to describe 
$\rho(T)$ above room temperature, incorporating the electrical transport in tunneling conduction junctions and intra-grain areas and inter-grain areas within  MLG/UG islands. The $\rho(T)$ of our GF samples investigated, as well as similar GF samples reported in the literature, exhibit good consistency between experimental data and the fits to the conduction network model, indicating the reliability and universality of the model in GF-based conduction systems. With the assistance of the conduction network model, we found that the temperature dependence (either insulating or metallic) in resistivity of pristine GFs is mainly determined by the discrete quantities of tunnel junctions and MLG/UG islands, and could be modulated by the combined effect of local strain-induced phonon hardening, local strain-induced transport gap modulation, and alterations in conduction networks, all of which arise from pre-stretching. Our work provides guidelines for the interpretation and analysis of resistivities in GF-related systems, considering the interplay of transport in tunneling conduction junctions and intra-grain and inter-grain areas of MLG/UG islands.

\begin{acknowledgments}

This work is supported by the Research Center for Nature-Inspired Science and Engineering, The Hong Kong Polytechnic University under Work Progamme CE1T. The authors appreciate the fruitful discussions with Prof Xi Zhu of The Chinese University of Hong Kong, Shenzhen.

\end{acknowledgments}

\section*{Supplementary notes}
Supplementary Notes S1--S4: Analysis of Raman spectra and estimation of graphene grain size, conversion from Raman shift to local strain, supplemental data analysis, analytical expression for strain modulated momentum $\boldsymbol{k}$.

\bibliography{0Main}

\end{document}